\documentclass{jpsj2}
\title{
Phase diagram of $S=1$ XXZ chain with next-nearest-neighbor interaction
}

\author{\textsc{Takahiro Murashima} $^{1}$, 
\textsc{Keigo Hijii} $^{1}$, 
\textsc{Kiyohide Nomura} $^{1}$
and 
\textsc{Takashi Tonegawa}$^{2}$}

\inst{$^{1}$Department of Physics, Kyushu University, 
Fukuoka 812-8581, Japan\\
$^{2}$Department of Mechanical Engineering, Fukui University of
Technology, Fukui 910-8505, Japan
\\
}

\abst{The one dimensional $S=1$ XXZ model with
next-nearest-neighbor interaction $\alpha$ 
and Ising-type anisotropy $\Delta$ 
is studied by using a numerical diagonalization technique.
We discuss the ground state phase diagram of this model 
numerically by 
the twisted-boundary-condition level spectroscopy method
and
the phenomenological renormalization group method,
and 
analytically by 
the spin wave theory.
We determine the phase boundaries among the XY phase, the Haldane phase, 
the ferromagnetic phase and the N\'{e}el phase, 
and then we confirm the universality class.
Moreover,
we 
map this model onto the non-linear $\sigma$ model 
and
analyze the phase diagram
in the $\alpha \ll -1$ and $\Delta \sim 1$ region
by using the renormalization group method.
}

\kword{spin-1, ground state phase diagram, frustration,
diagonalization,
level spectroscopy,
non-linear $\sigma$ model}

\begin{document}
\maketitle

\section{Introduction}
One dimensional antiferromagnetic Heisenberg
spin systems have been an important subject 
in many-body physics for a long time.
There are no ordered ground states 
with a continuous symmetry breaking 
in these systems, 
because of quantum fluctuation 
and low dimensionality.
It is well known 
from Haldane's conjecture\cite{Haldane1983,Haldane1983-b}
that there is a fundamental 
difference between half-integer 
and integer spin chains in their low-lying excitations.
In some lattice systems,
frustration, which is geometrically represented 
by a triangular lattice, tends to disorder the ground state.
We expect that
competing systems between quantum fluctuation 
and frustration
plus low dimensionality
exhibit fascinating phenomena in their ground states 
and low-lying excitations.
Recently, the materials of the $S=1$ spin chain
with next-nearest-neighbor interaction 
have been reported experimentally 
(CaV$_2$O$_4$\cite{KikuchiChibaKubo2001}, 
NaV(WO$_4$)$_2$\cite{MasudaSakaguchiUchinokura2002}).

In this paper, we study the $S=1$ XXZ chain with
next-nearest-neighbor (NNN) interaction,
which is described by the following 
Hamiltonian:
\begin{eqnarray}
{\cal H}
&=&\sum_{i=1}^{L} (S_i^xS_{i+1}^x + S_i^yS_{i+1}^y + 
\Delta S_i^z S_{i+1}^z)\nonumber\\
&&+ \alpha \sum_{i=1}^{L} (S_i^xS_{i+2}^x + S_i^yS_{i+2}^y + 
\Delta S_i^z S_{i+2}^z)\, ,\label{Hamiltonian}
\end{eqnarray}
where $\alpha$ is next-nearest-neighbor interaction constant
and $L$ is system size ($L$: even).
Hereafter, periodic boundary condition (PBC) 
will be assumed unless specifically
mentioned.

A phase diagram of 
this model
has been studied by the exact diagonalization method
for the $|\Delta| < 2 $ and
$0 \le \alpha < 0.3 $ region\cite{Tonegawa-Suzuki-Kaburagi},
and also by 
the infinite-system
density matrix renormalization group method (DMRG) for 
the $\Delta \ge 0$
and $\alpha \ge 0$ region
\cite{Kaburagi-K-H,Hikihara-K-K-T}. 
Many phases of the ground state have been found there;
the Haldane phase, 
the double Haldane phase, 
the N\'{e}el phase, 
the double N\'{e}el phase,
the gapped chiral phase 
and the gapless chiral phase.
The features of these phases are explained simply as follows.
The Haldane state has an energy gap,
where the spin correlation
decreases exponentially 
as $L \to \infty$.
The N\'{e}el state is 
characterized by
twofold degenerate ground states,
where
the staggered magnetization exhibits a long range order.
In the double Haldane phase, the system can be regarded as 
two Haldane subchains because of competition between
the strong next-nearest-neighbor
interaction 
and the weak nearest-neighbor interaction.
A first-order phase transition occurs
between the Haldane phase and the double Haldane phase
\cite{KolezhukRothSchollwock1996,KolezhukRothSchollwock1997}. 
We can also regard the double N\'{e}el state as two N\'{e}el
subchains (up-down-up-down or up-up-down-down).
The gapped chiral state and the gapless chiral state have
a chiral long range order, 
whereas the two spin correlation decays 
exponentially in the former state
and algebraically in the latter one
\cite{Kaburagi-K-H,Hikihara-K-K-T}.

On the $\alpha=0$ line, it is known that 
 four different ground
states exist;
the ferromagnetic state($\Delta \le -1$), 
the XY state($-1 \le \Delta \le 0$), 
the Haldane state($0 \le \Delta \lesssim 1.17$) and 
the N\'{e}el state($1.17 \gtrsim \Delta$)
\cite{Nomura1989,KitazawaNomuraOkamoto1996,KitazawaNomura1997,ChenHidaSanctuary2003}.
The ferromagnetic state has a long range order of the magnetization.
In the XY phase, the spin correlation shows an algebraic decay, and the
correlation length is infinite.
A first-order phase transition occurs
between the ferromagnetic phase and the XY phase,
and the 2D-Ising type phase transition occurs
between the Haldane phase and the N\'{e}el phase\cite{Nomura1989}.

It has been well known that 
the Berezinskii-Kosterlitz-Thouless (BKT) transition occurs
between the XY phase and the Haldane phase,
which is explained
by the bosonization method
for a spin-$s$ XXZ model\cite{Schulz1986}.
This transition belongs to the infinite-order phase transition 
in the traditional classification.
On this transition point, 
the logarithmic correction appears in
the correlation function, the critical exponents and so on.
Numerically, 
the BKT transition has been known to occur at the $\Delta = 0$ and 
$\alpha= 0$ point
in a spin-1 ($S=1$) XXZ model with bond
alternation constant $\delta$ for $|\delta|< 0.23$
\cite{KitazawaNomuraOkamoto1996,KitazawaNomura1997},
and that with uniaxial
single-ion-type anisotropy constant $D$
for $-2 < D < 0.5$
\cite{ChenHidaSanctuary2003}.
A bosonization study for a spin-$s$ XY model with NNN interaction
has supported that
this particular point is determined as the BKT transition point
in the $S=1$ case\cite{LecheminantJolicoeurAzaria}.
Recently, 
the fact that this point 
($\Delta = \alpha = 0$) is 
the BKT transition point
has been analytically proved\cite{Kitazawa2003}.
For many years, 
it was a difficult problem to determine numerically the critical point
and the universality class of the BKT transition, 
because of logarithmic corrections and the slow divergence 
of the correlation length.
Although the DMRG method is a powerful method to examine many phases,
it is not appropriate to determine the BKT transition line.
This problem was successfully resolved by 
the level spectroscopy method\cite{Nomura1995},
based on the conformal field theory 
and the renormalization group. 
Moreover, 
this method was improved 
by the twisted-boundary-condition 
level spectroscopy method\cite{NomuraKitazawa1998}.
We use the twisted-boundary-condition level spectroscopy method
to determine the BKT transition point when $\Delta \ne 0$ or 
$\alpha \ne 0$.

In this work, we present a phase diagram of the $S=1$ XXZ chain 
with NNN interaction,
analyzing the diagonalization data by 
the twisted-boundary-condition level spectroscopy method
\cite{NomuraKitazawa1998},
the phenomenological renormalization group (PRG)
method\cite{Barber,Roomany-Wyld}, 
the spin wave theory and
the renormalization group method.

This paper is organized as follows.
In the next section,
the obtained phase diagram is presented.
The numerical methods to determine the phase boundaries
are explained in \S \ref{NMAR}.
In \S \ref{ANLSM},
we map eq. (\ref{Hamiltonian}) onto the non-linear $\sigma$ model,
and then we 
analytically examine the phase diagram 
in the region of $\alpha \ll -1$ and $\Delta \sim 1$.
The final section is devoted to a summary and a discussion.

\section{Phase diagram}
\begin{figure}
\begin{center}
 \includegraphics[scale=0.5]{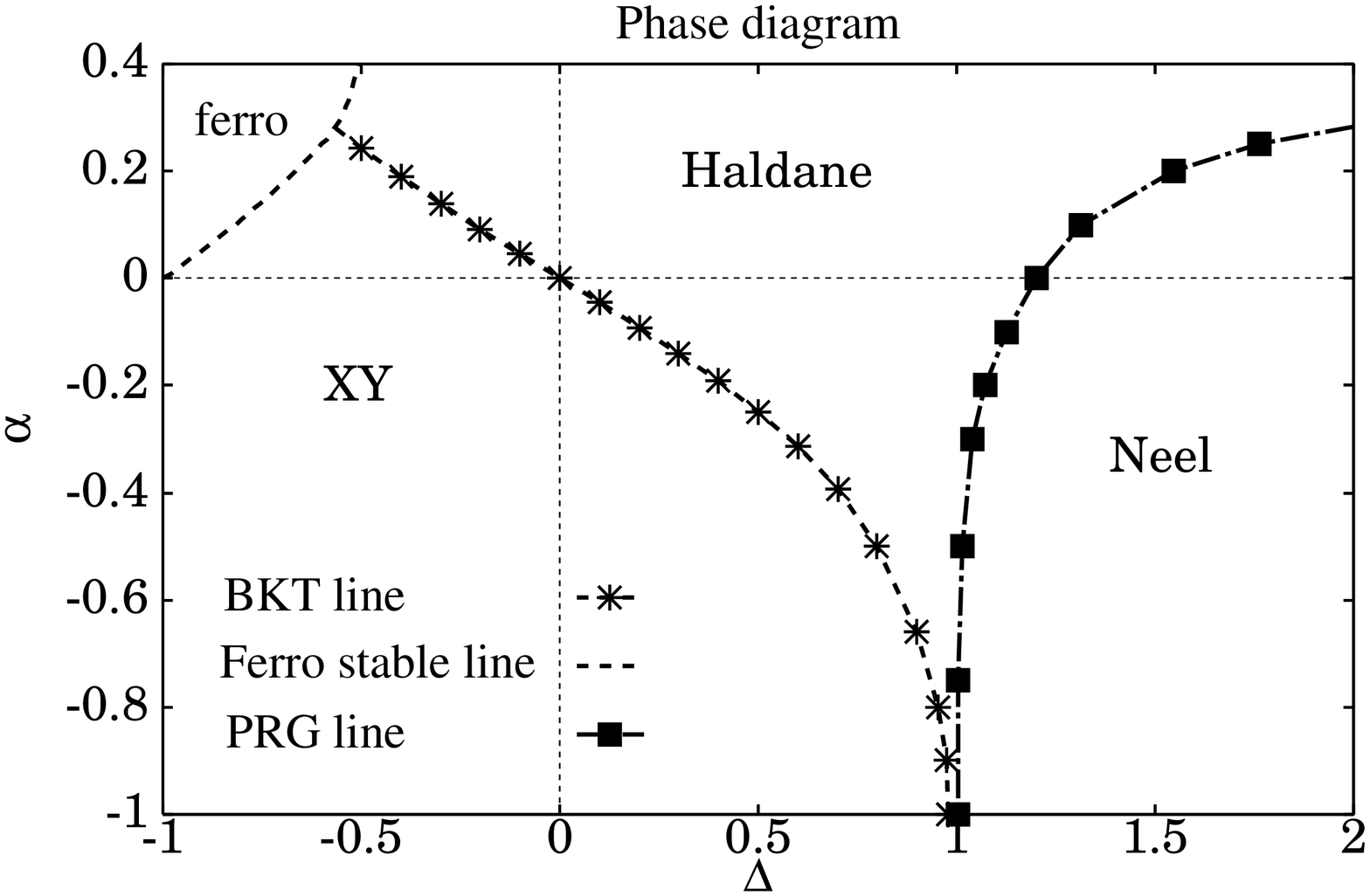}%
\end{center}
 \caption{Phase diagram of the $S=1$ XXZ chain with
  next-nearest-neighbor interaction in the
region of $-1 \le \Delta \le 2$ and $-1 \le \alpha \le 0.4$.
Note that the ferromagnetic stable line for $\alpha > 0.4$
is given in Fig. \ref{Ferro}.}
\label{phasediagram}
\end{figure}

The obtained phase diagram is summarized in Fig. \ref{phasediagram},
which
consists of the ferromagnetic
phase, the XY phase, the Haldane phase and the N\'{e}el phase.
We have determined the BKT transition line 
between the XY and Haldane phases
by using the twisted-boundary-condition 
level spectroscopy method\cite{NomuraKitazawa1998}. 
Then, from the analysis of the spin wave theory,
we determine the first-order phase transition line between
the ferromagnetic and XY phases, 
and that between the ferromagnetic and Haldane phases.
Moreover, the Haldane-N\'{e}el transition line is determined by 
the PRG method\cite{Barber,Roomany-Wyld}.

In the $\alpha > 0.4$ region,
it is expected that
the gapped and gapless chiral phases,
the double Haldane phase and the double N\'{e}el phase
appear in the phase diagram
\cite{KolezhukRothSchollwock1996,KolezhukRothSchollwock1997,
Kaburagi-K-H,Hikihara-K-K-T}.
Since the level spectroscopy and PRG methods do not work 
in this region because of incommensurability,
we do not treat this region in this study.
This point will be discussed further
in \S \ref{Discuss}.

\section{Numerical methods and results \label{NMAR}}

\subsection{Level spectroscopy and BKT transition \label{LS}}

In this subsection, we briefly review
the level spectroscopy method\cite{Nomura1995} 
and the twisted-boundary-condition
level spectroscopy method\cite{NomuraKitazawa1998},
and then we determine the BKT transition line
using the latter method.
In addition, we check the consistency of the universality class.

With the use of the bosonization technique\cite{Schulz1986},
the one dimensional XXZ spin system can be mapped onto the following
sine-Gordon model (in Euclidean space-time):
\begin{equation}
 S= \frac{1}{2\pi K}\int {\rm d}\tau {\rm d}x \left\{({\rm \partial}_{\tau} \phi)^2 +
 (\partial_x \phi)^2 \right\} + \frac{y_2}{2 \pi a^2}\int {\rm d}\tau
 {\rm d}x
 \cos{\sqrt{2}\phi}, \label{sineGordon}
\end{equation}
where $\phi \equiv \phi(\tau,x)$ is the boson field and 
$a$ is the lattice constant. 
After a scaling transformation $a \to {\rm e}^{\delta l} a$, we obtain the
following renormalization group equations:
\begin{equation}
 \frac{{\rm d} K^{-1}}{{\rm d}l} = \frac{1}{8} y_2^2,\quad
 \frac{{\rm d}y_2}{{\rm d}l}=\left(2-\frac{K}{2}\right)y_2.
\end{equation}
These equations show that 
the $y_2$ term
in eq. (\ref{sineGordon}) is an irrelevant field
for $K > 4$ and a relevant field for $K < 4$. 
On the $K=4$ line, 
the BKT transition occurs.
Denoting $K=4+2y_1$ near $K=4$, we have
\begin{eqnarray}
 \frac{{\rm d} y_1(l)}{{\rm d} l} &=&
  -y_2(l)^2 \label{K1},\\
 \frac{{\rm d} y_2(l)}{{\rm d} l} &=&
  -y_1(l)y_2(l). \label{K2}
\end{eqnarray}
The renormalization flow 
separatrix is given by $y_1(l)^2 - y_2(l)^2 = 0$ ($y_1(l)>0$).
These differential equations 
are the famous Kosterlitz equations\cite{Kosterlitz1974}.
On this separatrix,
the solution of eqs. (\ref{K1}) and (\ref{K2}) is
$y_2(l)=\pm y_1(l)=y_1/(1+y_1 l)$,
where $y_1$ is the bare coupling constant and $l=1/\ln L$.

When the $y_2$ term is relevant,
the scaling dimension $x$ should be corrected as $x\to x+Cy_2$.
Since the behavior of some scaling dimensions is different
depending on whether 
in the case of relevant $y_2$ or 
in the case of irrelevant $y_2$,
we expect that 
the BKT transition point can be determined 
by examining the scaling dimension.

Now,
we introduce the dual field $\theta \equiv \theta(\tau,x)$ to 
the boson field $\phi$
by the following equations:
\begin{equation}
 \partial_{\tau} \phi = -{\rm i} K \partial_x \theta, \quad
\partial_x \phi = {\rm i} K \partial_{\tau} \theta.
\end{equation}
The vertex operators are expressed in terms of 
$\theta$ and
$\phi$
as
\begin{equation}
 V_{n,m} = \exp({\rm i} n \sqrt{2}\theta) \exp({\rm i} m \sqrt{2}\phi), 
\label{VO}
\end{equation}
where $n$ and $m$ are integers,
conventionally called the electric charge and the magnetic charge,
respectively. 
Then, for the free field case ($y_2 = 0$),
the scaling dimensions of the vertex operators 
$V_{n,m}$
are\cite{KadanoffBrown1979}
\begin{equation}
 x_{n,m} 
= \frac{1}{2}\left(\frac{n^2}{K} + K m^2\right).\label{scalingdimension}
\end{equation}
In addition, there is a marginal operator 
${\cal M}=(\partial_{\tau}\phi )^2 + (\partial_x \phi)^2$,
whose scaling dimension is given by $x_{\textrm{marg}}=2$. 
From the conformal field theory, the eigenvalues $E_{n,m}$ with 
PBC are related to the scaling dimensions $x_{n,m}$ as\cite{Cardy1984}
\begin{equation}
 E_{n,m}(L) - E_{\textrm g}(L) = \frac{2\pi v_{\textrm s} x_{n,m}}{L} \label{energygap}
\end{equation}
in the limit of $L \to \infty$, where $E_{\textrm g}(L)$ is 
the ground state energy for the size $L$ system under PBC, and
$v_{\textrm s}$ is 
the spin wave velocity.
Using eqs. (\ref{scalingdimension}) and (\ref{energygap}),
we obtain
\begin{equation}
x_{\pm 4,0} = x_{0,\pm 1}=x_{\textrm{marg}} \label{orgx}
\end{equation}
when $K=4$ and $y_2=0$ (marginal).
Considering the invariance under $\phi \leftrightarrow -\phi$,
we obtain the combined operators
$\cos(\sqrt{2}\phi)\sim V_{0,+1} + V_{0,-1}$ 
and $\sin(\sqrt{2}\phi)\sim V_{0,+1} - V_{0,-1}$,
whose scaling dimensions are $x_{0,\cos}$
and $x_{0,\sin}$, respectively.
In the case of $y_2 \ne 0$,
these scaling dimensions 
differ from each other
because of the $y_2 \cos(\sqrt{2}\phi)$ term in eq. (\ref{sineGordon}).
Thus,
$x_{\pm 4,0}$, $x_{0,\cos}$, $x_{0,\sin}$ and $x_{\textrm{marg}}$
are expressed with
the logarithmic correction as
$x_{\pm 4,0}=2-y_1(l)$, $x_{0,\cos}=2+2y_1(l)(1+2t/3)$,
$x_{0,\sin}=2+y_1(l)$ and
$x_{\textrm{marg}}=2-y_1(l)(1+4t/3)$, where $t$ is
a distance from the BKT transition point, defined as
$y_2(l)=\pm y_1(l)(1+t)$.
At the BKT transition point ($t=0$),
we have
\begin{equation}
x_{\pm 4,0}=x_{\textrm{marg}}.
\end{equation}
Thus we can use this energy level crossing to determine the
BKT transition point\cite{Nomura1995}.

Although the above procedure 
can be applied to determine the BKT transition point,
a better method,
the twisted-boundary-condition level spectroscopy method,
has been proposed\cite{NomuraKitazawa1998}. 
This method is based on the $SU(2)/Z_2$
symmetry on the BKT transition point.
We use the following twisted boundary condition (TBC):
\begin{equation}
 S_{L + j}^{x,y} = -S_j^{x,y}, S_{L + j}^z = S_{j}^z.
\end{equation}
Then, the scaling dimensions of the
eigenvalues $E_{n,m}^{\textrm{TBC}}(L)$ with TBC are given by
\begin{equation}
 x_{n,m}^{\textrm{TBC}} = 
\frac{1}{2}\left\{\frac{n^2}{K} + K\left(m +
 \frac{1}{2}\right)^2\right\} 
= x_{n,m+1/2} \label{xTBC}
\end{equation}
or
\begin{equation}
 E_{n,m}^{\textrm{TBC}}(L)-E_{\textrm g}(L)=\frac{2\pi v_{\textrm
  s}}{L}x_{n,m}^{\textrm{TBC}}.
\end{equation}
Since
we can obtain the half-integer magnetic
charge effectively from eq. (\ref{xTBC}),
we now extend the magnetic charge $m$ to that including half-integers.
From eqs. (\ref{scalingdimension})
and (\ref{xTBC}),
we have in the case of $y_2=0$,
\begin{equation}
x_{\pm 2,0} = x_{0,0}^{\textrm{TBC}} (= x_{0,\pm 1/2}) \label{BKT}
\end{equation}
because $K=4$.
When $y_2 \ne 0$,
following the above mentioned way,
we define the scaling dimensions
$x_{0,\cos}^{\textrm{TBC}}$ and $x_{0,\sin}^{\textrm{TBC}}$
of the operators
$\cos(\phi/\sqrt{2}) \sim V_{0,+1/2} + V_{0,-1/2}$
and $\sin(\phi/\sqrt{2}) \sim V_{0,+1/2} - V_{0,-1/2}$, respectively,
and also we express these scaling dimensions with the logarithmic
correction as shown in Table \ref{BKTtable}.
At the BKT transition point ($y_2=\pm y_1$, $y_1>0$),
we have
\begin{equation}
x_{\pm 2,0}=x_{0,\sin}^{\textrm{TBC}}.
\end{equation}
Since 
the electric charge and the magnetic charge of 
$x_{\pm 2,0}$ and $x_{0,\sin}^{\textrm{TBC}}$
are smaller than those of $x_{\pm 4,0}$ and $x_{\textrm{marg}}$,
we can obtain more accurate results by using
the twisted-boundary-condition
level spectroscopy
 method\cite{NomuraKitazawa1998} 
than those by using
the original level spectroscopy
method\cite{Nomura1995}.

\begin{table}
\caption[x]{This table shows relations between quantum numbers 
$(M,P,k)$ and renormalized scaling dimensions $x$. $M=\sum_{i=1}^L S_i^z$.
BC represents boundary conditions. 
PBC means periodic boundary condition 
and TBC means twisted boundary condition. Under TBC, 
the parity is different from that of PBC, 
so we use $^*$ to classify.
$y_2(l)=\pm y_1(l)=y_1/(1+y_1 l)$, where
$y_1$ is the bare coupling constant and $l=\ln L$.
} 

\begin{center}
\begin{tabular*}{150mm}[t]{@{\extracolsep{\fill}}ccccc||cc}

\hline
$M$ & BC   & $P$  &$k$      
&correction in $x$&operator in s.G.&scaling dimension\\ 
\hline
$\pm$2  & PBC  & 1  &0      
&$1/2 - y_1(l)/4$&$\exp(\pm {\rm i} 2 \sqrt{2}\theta)$&$x_{\pm2,0}$\\ 
0       & TBC  & -1$^*$ &   
&$1/2+y_1(l)/4-y_2(l)/2$&$\sin(\phi / \sqrt{2})$&$x_{0,\sin}^{\textrm{TBC}}$\\ 
0       & TBC  & 1$^*$  &   
&$1/2+y_1(l)/4+y_2(l)/2$&$\cos(\phi / \sqrt{2})$&$x_{0,\cos}^{\textrm{TBC}}$\\
\hline
\label{BKTtable}
\end{tabular*}
\end{center}
\end{table}

\begin{figure}
\begin{center}
\includegraphics[scale=0.5]{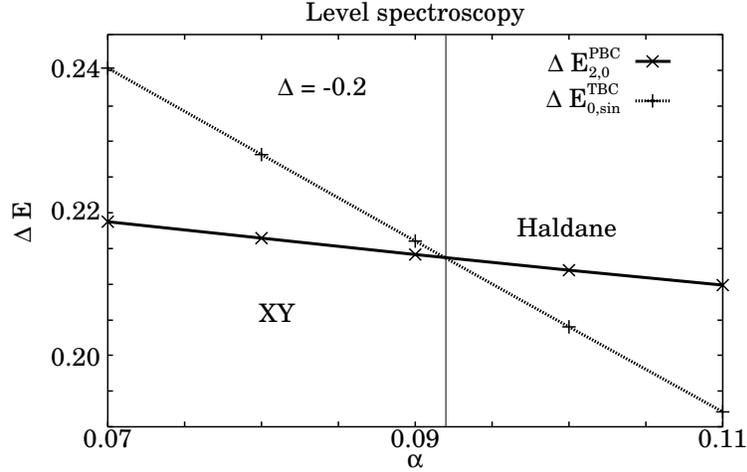}
\end{center}
\caption{Energy level crossing 
between $x_{2,0}$ and 
$x_{0,\sin}^{\textrm{TBC}}$
when $L=16$ and $\Delta = -0.2$. 
The value of $\alpha_{\rm c}^{\rm{BKT}}$ is obtained to be
$\alpha_{\textrm c}^{\rm{BKT}} = 0.09193$.}
\label{BKTcross}
\end{figure}

We can identify the correspondence between the operators in the boson
representation and the eigenstates of spin chains by comparing their
symmetry properties\cite{NomuraKitazawa1998} 
as shown in Table \ref{BKTtable}. 
The Hamiltonian of eq. (\ref{Hamiltonian}) with PBC is 
invariant under the spin rotation around the z-axis,
the translation (${\bf S}_i \to {\bf S}_{i+1}$) and
the space inversion (${\bf S}_i \to {\bf S}_{N-i+1}$). 
The quantum numbers corresponding to these invariance
are $M \equiv \sum_{i=1}^L S_i^z$,
$k=2\pi n/L (n=0,\cdots,L/2-1)$ and 
$P=\pm 1$, respectively.
From Table \ref{BKTtable}, 
we see that
$x_{\pm 2,0}$ corresponds to 
the lowest excitation energy with $M=\pm 2$, $P=1$, $k=0$, 
and
$x_{0,\sin}^{\textrm{TBC}}$ to the lowest excitation energy with $M=0$, $P=-1^*$ 
under TBC.
The energy level crossing of these two energies 
for $L=16$ and $\Delta=-0.2$
is shown in 
Fig. \ref{BKTcross}.

\begin{figure}
  \begin{center}
 \includegraphics[scale=0.5]{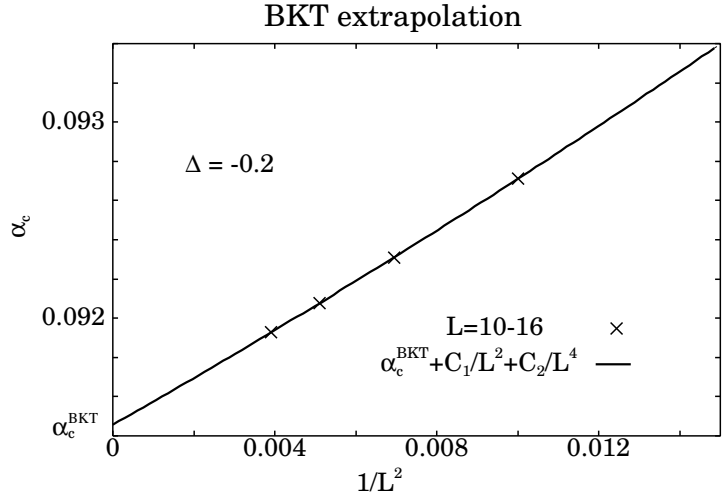}
   \end{center}
 \caption{Extrapolation procedure of $\alpha_{\textrm c}^{\rm{BKT}}$
 for $\Delta=-0.2$,
where a fitting function 
$\alpha(L)=\alpha_{\textrm c}^{\rm{BKT}} + C_1/L^2 + C_2/L^4$,
is assumed.
}
\label{BKTextrapolation}
 \end{figure}

In addition to the logarithmic correction,
there is another correction 
between a lattice system and a continuous model.
Considering the $x=4$ irrelevant field operator 
$L_{-2}\bar L_{-2}\bf{1}$ 
in terms of the conformal field theory\cite{Cardy1986NB},
we should extrapolate the infinite-size critical point 
$\alpha_{\rm c}^{\rm{BKT}}$ as
\begin{equation}
\alpha(L) = \alpha_{\textrm c}^{\rm{BKT}} +C_1/L^2 + C_2/L^4. 
\end{equation}
The BKT transition point for $\Delta = -0.2$
is determined as shown in Fig.
\ref{BKTextrapolation}. 
The same procedure can be carried out 
varying $\alpha$ with fixed $\Delta$,
and the obtained BKT transition line is represented in Fig. 
\ref{phasediagram}.

Next, in order to check the consistency of the universality class, 
we confirm that two universal relations,
concerning
the central charge $c$ and the energy gap ratio,
hold in the following way.

\begin{figure}
\begin{center}
\includegraphics[scale=0.5]{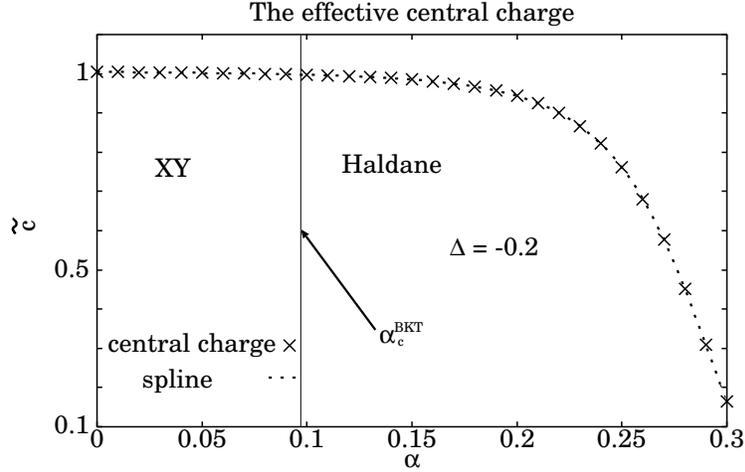}%
\end{center}
\caption{Plot of the effective central charge $\tilde{c}$ 
as a function of $\alpha$ for $\Delta=-0.2$.
In the massless region, $\tilde{c}=1$.
The BKT transition occurs at $\alpha_{\textrm c}^{\rm{BKT}}$.}
\label{central}
\end{figure}

The BKT transition line is expected to be described by the conformal
field theory with the central charge $c=1$. 
It is well known that the finite-size correction to the ground state
energy is related to 
the central charge $c$ and the spin wave velocity
$v_{\textrm s}$
as follows\cite{Cardy1984,BloteCardyNightingale1986,
Affleck1986,Cardy1986NB},
\begin{equation}
 \frac{1}{L}E_{\textrm g}(L)\cong \varepsilon_{\infty} - 
\frac{\pi c v_{\textrm s}}{6 L^2}, \label{centraleq}
\end{equation}
where 
\begin{equation}
 v_{\textrm s} = \lim_{L \to \infty}
  \frac{L}{2\pi}\{E_{k_1}(L)-E_{\textrm g}(L)\}.
\end{equation}
Here $E_{k_1}(L)$ is the
energy of the excited state with wave number $k_1 = \frac{2\pi}{L}$ and
magnetization $M = 0$.
In general, we should obtain
$v_{\textrm s}$ as
\begin{equation}
 v(L) = v_{\textrm s} + C_1/L^2 + C_2/L^4.
\end{equation}
Although there remain logarithmic corrections in the central
charge, 
which is obtained from eq. (\ref{centraleq}),
they are small enough to be $O(1/(\ln L)^3)$\cite{Cardy1986}.
Thus we neglect them. 
It is proper that 
the effective central charge $\tilde{c}$ 
should
correspond to the 
central charge $c$ in a massless region.
In a massive region, 
the correction term in
eq. (\ref{centraleq}) is modified as $1/L^2 \to \exp(-L/\xi)$.
We calculate the effective central charge $\tilde{c}$, 
the result of which is shown
in Fig. \ref{central}.
We can see that the effective central charge changes rapidly 
from $\tilde{c}=1$ (massless) to $\tilde{c}=0$ (massive).
This fact agrees with 
previous results in
refs. \citen{InoueNomura1999} and \citen{Inoue2000}.

\begin{figure}
\begin{center}
\includegraphics[scale=0.5]{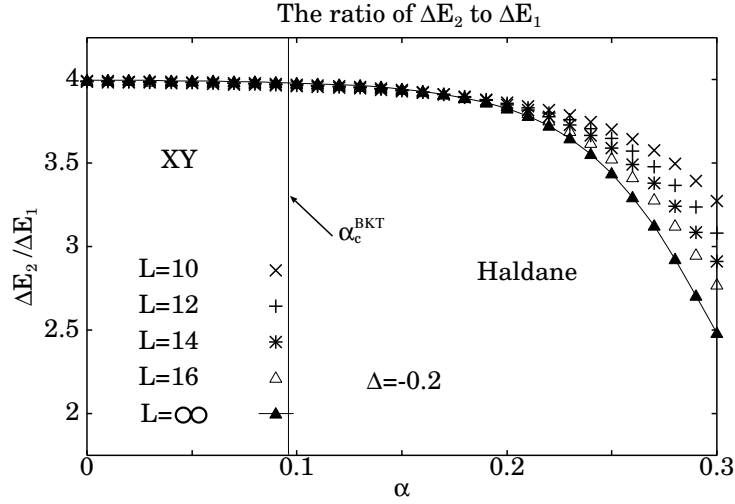}
\end{center}
\caption{Plot of
$\Delta E_2/\Delta E_1$
as a function of $\alpha$
for $\Delta=-0.2$.
In the XY phase, $\Delta E_2/\Delta E_1=4$, 
and
in the gapful phase,
$\Delta E_2/\Delta E_1=2$.
Because of the finite-size effect, 
$\Delta E_2/\Delta E_1$ varies slowly.
}
\label{universal4}
\end{figure}

Next,
we examine the ratio of 
the energy levels.
From eqs. (\ref{scalingdimension}) and (\ref{energygap}),
we obtain
\begin{equation}
 \frac{\Delta E_2}{\Delta E_1} 
= 4, \label{univ4}
\end{equation}
where $\Delta E_i = E(M=\pm i) - E_{\textrm g}(M=0)$ ($i=1,2$).
This relation is independent of $K$. 
In the gapful (massive) phase, on the other hand,
we have
\begin{equation}
 \frac{\Delta E_2}{\Delta E_1} 
\to 2 \quad (L \to \infty), \label{univ2}
\end{equation}
since the mass of two magnons
is twice as large as
that of one magnon when each magnons are independent.
This ratio varying with $\alpha$
in the case of $\Delta=-0.2$
is shown in Fig. \ref{universal4}.
We can also see the size dependence of $\Delta E_2/\Delta E_1$.
Affected by the $x=4$ irrelevant field operator $L_{-2}\bar L_{-2}\bf{1}$,
the extrapolated value,
described in Fig. \ref{universal4}, 
can be given
as follows,
\begin{equation}
{\cal O} = {\cal O}_{\infty} + C_1/L^2 + C_2/L^4,
\end{equation}
where ${\cal O}$ means $\Delta E_2/ \Delta E_1$.
It appears that the ratio is 4 in the massless phase 
and it changes very slowly from 4 to 2
because of the finite-size effect.

\subsection{Instability of ferromagnetic phase}

We determine the unstable boundary of the ferromagnetic phase as follows.
If there is no intermediate phase,
we can obtain the boundary of the ferromagnetic phase 
by examining the energy level crossing
between
the energy of the ferromagnetic state, $E = (1 +
\alpha)\Delta L$, and that of the $M = 0$ state.
Moreover, we compare the numerical result with 
that obtained by
the ferromagnetic spin wave theory
(see APPENDIX \ref{app}).

\begin{figure}
\begin{center}
\includegraphics[scale=0.5]{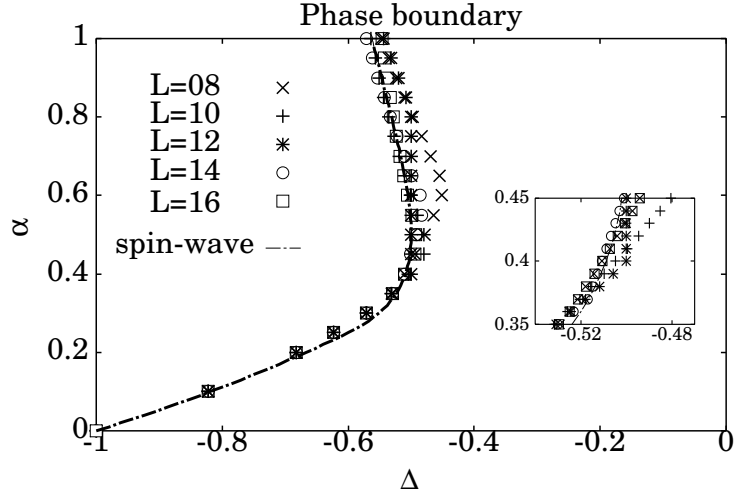}
\end{center}
\caption{Phase boundary of the ferromagnetic
 phase, obtained by the energy level crossing from 
the diagonalization data and the spin wave theory.
}
\label{Ferro}
\end{figure}

The phase diagram 
near the ferromagnetic phase determined by these two methods
is shown in Fig. \ref{Ferro}.
Below $\alpha = 0.38$, the ferromagnetic stable line obtained from 
the diagonalization data agrees with the line determined by 
the spin wave theory.
The fact that the convergence of the ferromagnetic boundary 
depending on $L$
is not so good above $\alpha = 0.38$ suggests
that the incommensurability begins 
to modulate the ground state energy.
The behavior of the ferromagnetic phase boundary
obtained by
the energy level crossing
is,
however,
analogous to that by the spin wave theory.
Therefore,
we adopt the line of the spin wave theory
as the ferromagnetic phase boundary in Fig. \ref{phasediagram}.

\subsection{Haldane-N\'{e}el transition}
\begin{figure}
\begin{center}
 \includegraphics[scale=0.5]{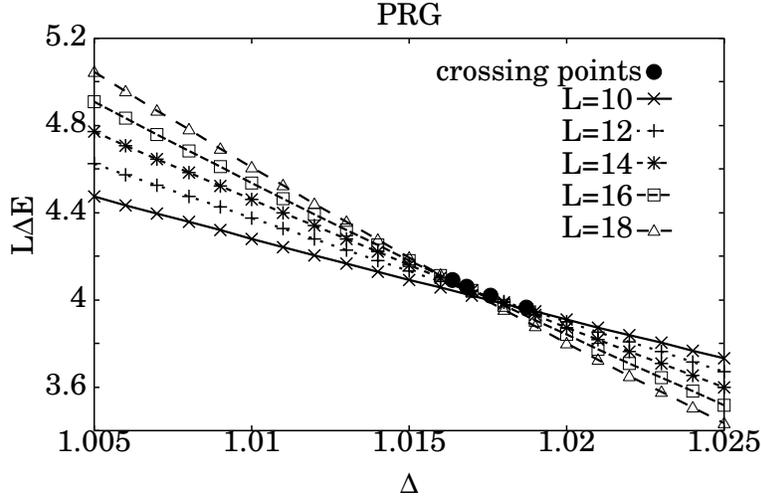}%
\end{center}
\caption{The $\Delta$ dependence of $L\Delta E(L)$ with $\alpha=-0.5$
for $L=$10, 12, 14, 16 and 18. The crossing points ($\bullet$) 
are the finite-size transition points.}
\label{prg}
 \end{figure}

\begin{figure}
\begin{center}
 \includegraphics[scale=0.5]{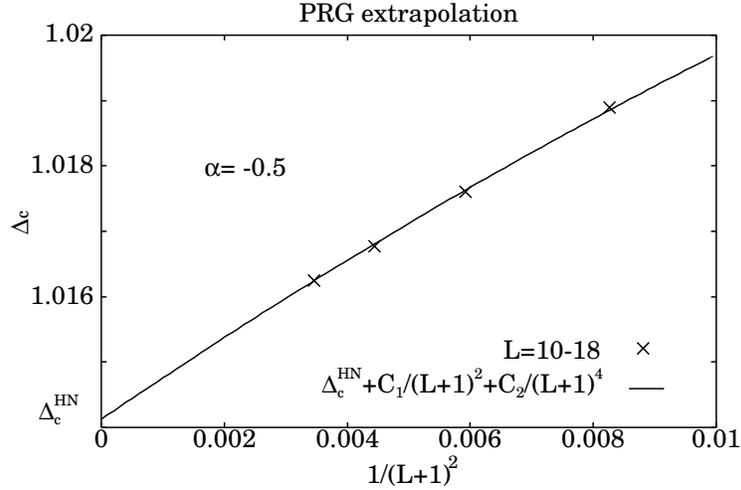}%
\end{center}
 \caption{Extrapolation procedure of $\Delta_{\rm c}^{\rm{HN}}$
for $\alpha=-0.5$,
where a fitting function 
$\Delta_{\rm c}(L,L+2)=\Delta_{\textrm c}^{\textrm{HN}}
+C_1/(L+1)^2+C_2/(L+1)^4$ is assumed.}
\label{prginfty}
 \end{figure}

\begin{figure}
\begin{center}
 \includegraphics[scale=0.5]{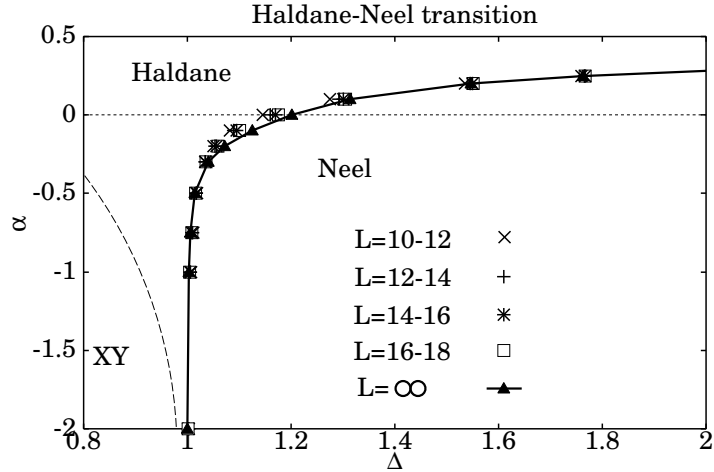}
\end{center}
 \caption{Phase diagram near the Haldane-N\'{e}el transition line.}
\label{HalNeel}
\end{figure}

The phase transition between the Haldane phase and 
the N\'{e}el phase
belongs to the second order phase transition.
In order to determine the transition line between these phases, 
we apply the PRG method\cite{Barber,Roomany-Wyld}.
This method is explained as follows.
In the N\'{e}el phase, the ground state is twofold degenerate 
in the thermodynamic limit. 
For finite $L$, however, this degeneracy is splitting 
into two energy levels.
The energy gap between these energy levels is
$\Delta E(L,\alpha,\Delta)=E(M=0,P=-1,k=\pi)-E_{\textrm g}(M=0,P=1,k=0)$, 
which behaves as
$\Delta E \sim \exp(-L/\xi)$
in the $L \to \infty$ limit.
On the other hand, the Haldane state has the energy gap,
which remains finite in the thermodynamic limit.
Thus the product $L\Delta E(L)$ increases (decreases) with $L$
for 
$\Delta < \Delta_{c}^{\textrm{HN}}$ ($\Delta > \Delta_{c}^{\textrm{HN}}$),
where $\Delta_{c}^{\textrm{HN}}$ means the critical value of $\Delta$ at
the Haldane-N\'{e}el transition.
Therefore, we use the following equation to determine 
the Haldane-N\'{e}el transition:
\begin{equation}
L\Delta E(L,\alpha, \Delta_{\textrm c}(L, L+2))=
(L+2)\Delta E(L+2, \alpha, \Delta_{\textrm c}(L, L+2)).
\end{equation}
The $\Delta$-dependences of $L\Delta E(L)$ with $\alpha = -0.5$
for various $L$ are
shown in Fig. \ref{prg}.
We extrapolate the infinite-size critical point 
$\Delta_{\textrm c}^{\textrm{HN}}$
as follows:
\begin{equation}
\Delta_{\textrm c}(L,L+2) = \Delta_{\textrm c}^{\textrm{HN}} + \frac{C_1}{(L+1)^2} 
+ \frac{C_2}{(L+1)^4}.
\end{equation}
When $\alpha = -0.5$,
the infinite-size transition point $\Delta_{\textrm c}^{\textrm{HN}}$ is determined as
shown in Fig. \ref{prginfty}.
The same procedure can be carried out by varying $\Delta$
with fixed $\alpha$.
The obtained Haldane-N\'{e}el transition line is shown 
in Fig. \ref{phasediagram} and also in Fig. \ref{HalNeel}
together with the finite-size results.

From the quantum Monte Carlo simulation in the case of $\alpha = 0$
\cite{Nomura1989},
this transition has been known 
to belong to the 2D-Ising type universality class.
Therefore, we expect that the universality class of the whole 
Haldane-N\'{e}el transition line
is of the 2D-Ising type.

\section{The analysis of the anisotropic non-linear 
$\sigma$ model \label{ANLSM}}

\begin{figure}
\begin{center}
\includegraphics[scale=0.5]{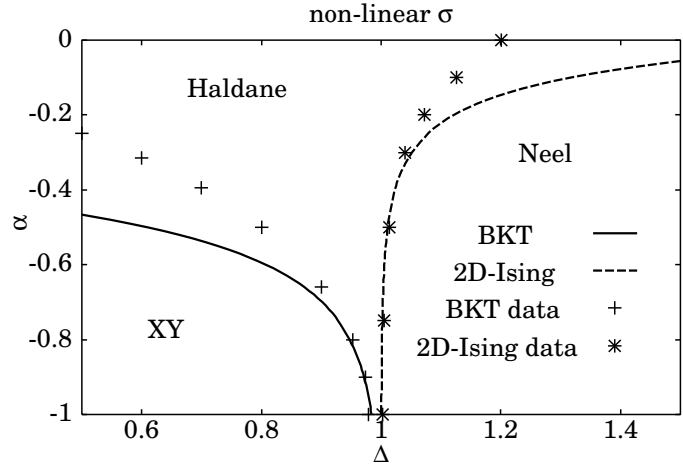}
\end{center}
\caption{Phase diagram 
near the $\alpha = -1$ and $\Delta =1$ point. 
We compare the schematic phase diagram 
obtained from eq. (\ref{nelpel1})
with 
the numerically extrapolated data to $L \to \infty$. 
The numerical results
 are in good agreement with the analytical line near $\Delta = 1$ and
$\alpha = -1$.}
\label{sigma2}
\end{figure}

In order to examine whether the Haldane phase does or does not survive
in the region of $\alpha \ll -1$ and $\Delta \sim 1$,
we perform the following analytical calculation, 
the details of which is discussed in APPENDIX \ref{app2}.
First, we map 
the spin-$s$ ($s$ is integer) XXZ spin chain with NNN interaction
onto the non-linear $\sigma$ model 
 with Ising-type anisotropy $\Delta$ in the large-$s$ limit 
\cite{Affleck1,Affleck2,Affleck-Les}.
Then, 
we obtain the renormalization group equation (eq. (\ref{RGeq}))
from the non-linear $\sigma$ model
by using the renormalization group method\cite{Nelson-Pelcovits}.
From this equation,
the renormalization group flows are represented 
by $\Delta$ and $\alpha$ as
\begin{equation}
|\Delta -1| \propto \frac{\sqrt{1-4\alpha}}{1-\alpha}
\exp{(-2\pi \sqrt{1-4\alpha})}, \quad
\text{when} \quad |\Delta -1| \ll 1. \label{nelpel1}
\end{equation}
We assume that these renormalization group flows are connected to 
each renormalization group fixed point:
the XY or the Ising fixed point.
These fixed points may be asymmetric on the phase diagram.
Therefore, we can draw the schematic phase diagram,
comparing with the numerical data,
as shown in Fig. \ref{sigma2}.
The numerical results
are in good agreement with the analytical lines 
near $\Delta \sim 1$ and $\alpha \sim -1$.

The phase is massive
at the isotropic point, $\Delta = 1$ and $\alpha \to -\infty$,
since the mass term (eq. (\ref{mass})) exists.
Thus, we may conclude that 
the phase transition between
the XY phase and the N\'{e}el phase does not directly occur 
in this limit
and the Haldane phase appears between these two phases.

\section{SUMMARY AND DISCUSSION} \label{Discuss}

The ground state phase diagram of the $S=1$ XXZ chain 
with NNN interaction has been determined numerically 
by analyzing the numerical diagonalization data using
the twisted-boundary-condition
level spectroscopy method\cite{NomuraKitazawa1998} and
the phenomenological renormalization group method
\cite{Barber,Roomany-Wyld},
and analytically by
the spin wave theory.
Particularly,
it has been resolved that the phase transition 
between the XY phase and the N\'{e}el phase 
does not occur directly for $\alpha \ll -1$ near $\Delta=1$
by mapping onto 
the non-linear $\sigma$ model\cite{Affleck1,Affleck2,Affleck-Les} and 
using the renormalization group method\cite{Nelson-Pelcovits}.

In this paper, we have not treated the region for $\alpha > 0.4$,
except for the vicinity of the ferromagnetic phase,
because of the incommensurability.
When $\Delta=1.0$, 
the commensurate-incommensurate (C-IC) change
is known to 
occur at $\alpha=\alpha_{\rm{C-IC}}$,
where $\alpha_{\rm{C-IC}} \sim 0.28$ is obtained by the DMRG method
\cite{KolezhukRothSchollwock1996,KolezhukRothSchollwock1997}
and 
$\alpha_{\rm{C-IC}} \sim 0.38$ by the diagonalization technique
\cite{Tonegawa-Kaburagi-Ichikawa-Harada}.
Note that this C-IC change 
is not a phase transition 
because the energy gap does not vanish.

In principle, 
the phase transition line between
the Haldane phase and the double Haldane phase 
can be determined
by Kitazawa's twisted-boundary-condition
level spectroscopy method\cite{Kitazawa1997}.
However,
we failed to determine this transition line
by this method.
Since the ground state energy 
is modulated with 
the long wave length in the incommensurate region,
the extrapolation procedure is difficult.
In this region, 
some long range orders, chiral order and string order, 
exist,
and therefore
the infinite-system DMRG method is known to be 
relatively useful
\cite{Kaburagi-K-H,Hikihara-K-K-T}.
However, it is difficult to determine  phase boundaries
of the gapless chiral phase,
where the string order parameter is
exhibiting a quasi-long-range order,
by using the DMRG method. 
In order to advance the analysis of the incommensurate region,
we need to investigate the incommensurability and we
must find an effective method to treat this nasty problem.

\section*{Acknowledgments}
The numerical calculation in this work is based on the program
packages TITPACK version 2, developed by Professor H. Nishimori,
and KOBEPACK/S, coded by Professor M. Kaburagi.  This work is
partly supported by a Grant-in-Aid for Scientific Research on
Priority Areas (B) ('Field-Induced New Quantum Phenomena in
Magnetic Systems') from the Ministry of Education, Culture,
Sports, Science and Technology of Japan.  We further thank
the Supercomputer Center, Institute for Solid State Physics,
University of Tokyo, the Information Synergy Center, Tohoku
University, and the Computer Room, Yukawa Institute for Theoretical
Physics, Kyoto University for computational facilities.

\appendix
\section{The analysis of spin wave instability\label{app}}

In this Appendix, we consider 
the instability analysis of the ferromagnetic phase
of the spin-$s$ XXZ chain 
with NNN interaction.
We consider the Hamiltonian of eq. (\ref{Hamiltonian}),
but $S^x,S^y$ and $S^z$
are spin-$s$ operators now.
When we denote the fully ferromagnetic state 
by $|0 \rangle$,
we obtain the ground state energy for the fully ferromagnetic phase
as 
\begin{equation}
{\cal H} |0 \rangle= 
E_0 |0\rangle,
\end{equation}
where
$E_0 =  \Delta s^2 L + \alpha \Delta s^2 L$
and $s$ is the eigen value of $S^z$ for $|0\rangle$.
Then, in order to see the lowest excitation,
we consider the following operator which is the linear combination of 
the spin lowering operator, 
\begin{equation}
\sum_m S^-_m \exp(- {\rm i} k m),
\end{equation}
where $k$ is the wave number.
When this operator acts on $|0 \rangle$,
we obtain 
\begin{equation}
{\cal H} \sum_m S^-_m \exp( - {\rm i} k m)
|0\rangle
=\{E_0 +\omega (k)\}
\sum_m S^-_m \exp( - {\rm i} k m)
|0\rangle,
\end{equation}
where $\omega(k)$ represents the dispersion relation,
\begin{eqnarray}
\omega(k)&=&2s \{ ( \cos k - \Delta )
+\alpha ( \cos 2k - \Delta    ) \} \\
&=&2s\left\{
2\alpha \left( \cos k + \frac{1}{4\alpha} \right)^2 
-\frac{1}{8\alpha}
-\alpha -\Delta - \alpha\Delta
\right\}. \label{omega}
\end{eqnarray}
If $\omega(k)$ is negative, the fully ferromagnetic state
is unstable because of the spin wave excitation.
Therefore
we can regard the line of $\omega(k)=0$
as the ferromagnetic stable-unstable boundary.
Since the prefactor $2s$ is positive,
we omit $2s$ in the following analysis.

From eq. (\ref{omega}),
we can see that the minimum value of the dispersion $|\omega(k)|$ 
is different whether in the case of
$\alpha > 1/4$ or in the case of $\alpha \le 1/4$.
In order to examine the incommensurate region,
it is enough to consider for $\alpha>0$.

In the former case,
$|\omega(k)|$ has a minimum value when $k=\pm \arccos\frac{-1}{4\alpha}$.
Then,
we have
\begin{equation}
\Delta=-\frac{1+8\alpha^2}{8\alpha \left( 1+\alpha\right)} . 
\label{dm05a05}
\end{equation}
Therefore the fully ferromagnetic phase becomes unstable for 
$k=\pm \arccos \frac{-1}{4\alpha}$ modes of the spin wave excitation.
Classically, these modes are connected with incommensurability.

In the latter case,
$|\omega(k)|$ is minimum 
when $k=\pm \pi$.
Then,
we have
\begin{equation}
\Delta=\frac{\alpha-1}{\alpha+1}.
\end{equation}
Therefore the fully ferromagnetic phase becomes unstable for 
$k=\pm \pi$ mode of the spin wave excitation.

\section{The non-linear $\sigma$ model \label{app2}}

The spin-$s$ XXZ chain can be mapped onto 
the non-linear $\sigma$ model with Ising-type anisotropy
in the large-$s$ limit\cite{Affleck1,Affleck2,Affleck-Les}.

We first consider the antiferromagnetic Heisenberg Hamiltonian with
NNN interaction $\alpha$:
\begin{equation}
{\cal H}=\sum_i{\bf S}_i \cdot {\bf S}_{i+1}
+\alpha\sum_i{\bf S}_i \cdot {\bf S}_{i+2}
\end{equation}
with $\alpha<0$. 
We want to keep low energy modes as follows,
\begin{equation}
 {\bf S}_i \simeq \left(-1\right)^i s\boldsymbol{\varphi}_i+a{\bf l}_i,
\end{equation}
where $\boldsymbol{\varphi}$ corresponds to the Fourier mode 
with the wave number $k\simeq\pi$, 
and ${\bf l}$ corresponds to the Fourier mode with the wave number 
$k\simeq 0$.
These operators are defined as
\begin{eqnarray}
\boldsymbol{\varphi}\left(2i+\frac{1}{2}\right)
&=&\frac{\left({\bf S}_{2i+1}-{\bf S}_{2i}\right)}{2s}, \\
{\bf l}\left(2i+\frac{1}{2}\right)
&=&\frac{\left({\bf S}_{2i+1}+{\bf S}_{2i}\right)}{2a} .
\end{eqnarray}
where $a$ is the lattice spacing.
Note that
$\boldsymbol{\varphi}$ and ${\bf l}$ are slowly varying 
function of $i$.
In the continuous limit,
these operators obey the following commutators,
\begin{eqnarray}
\left[l^\alpha(x),l^\beta(y)\right]&=&i\varepsilon^{\alpha\beta\gamma}
l^\gamma (x)\delta(x-y), \\
\left[l^\alpha(x),\varphi^\beta(y)\right]&=&i\varepsilon^{\alpha\beta\gamma}
\varphi^\gamma (x)\delta(x-y), \\
\left[\varphi^\alpha(x),\varphi^\beta(y)\right]&=&
i\varepsilon^{\alpha\beta\gamma}
l^\gamma (x)\delta(x-y) 2a \frac{1}{4s^2} \rightarrow 0,
\end{eqnarray}
where $\{\alpha, \beta, \gamma\} = \{x, y, z\}$
and $\varepsilon^{\alpha\beta\gamma}$ is the three-dimensional
Levi-Civita symbol.
Repeated indices are normally summed.
These operators satisfy the following constraints,
\begin{eqnarray}
\boldsymbol{\varphi} \cdot {\bf l}&=&0, \\
\left|\boldsymbol{\varphi}\right|^2&=&1.
\end{eqnarray}
Now, the Hamiltonian density can be mapped onto the $O(3)$ non-linear
$\sigma$ model,
\begin{equation}
{\cal H}=\frac{v}{2}\left\{
g\left({\bf l}-\frac{\theta}{4\pi}\frac{\partial \boldsymbol{\varphi}}
{\partial x}
\right)^2+\frac{1}{g}
\left(\frac{\partial \boldsymbol{\varphi}}{\partial x}\right)^2\right\},
\end{equation}
where the velocity $v=2sa$,
the topological angle $\theta=2\pi s$ and
the coupling constant $g=\frac{2}{s\sqrt{1-4\alpha}}$.

Next,
we consider the Hamiltonian with Ising-type anisotropy $\Delta$
as follows,
\begin{equation}
{\cal H}=\sum_i{\bf S}_i \cdot {\bf S}_{i+1}
+\alpha\sum_i{\bf S}_i \cdot {\bf S}_{i+2}
+\sum_i \left\{
 \left(\Delta-1\right) S^z_i S^z_{i+1} + \alpha \left(\Delta -1\right)
S^z_i S^z_{i+2}\right\}. \label{anisotropicspin}
\end{equation}
We only choose the leading term in the anisotropic terms as
\begin{eqnarray}
S^z_iS^z_{i+1}  &\sim&  -s^2\varphi_z^2, \\
S^z_iS^z_{i+2}  &\sim&  s^2\varphi_z^2 .
\end{eqnarray}
Then, we obtain the following effective Hamiltonian density,
\begin{equation}
{\cal H}=\frac{v}{2}\left\{
g\left({\bf l}-\frac{\theta}{4\pi}\frac{\partial \boldsymbol{\varphi}}
{\partial x}
\right)^2+\frac{1}{g}
\left(\frac{\partial \boldsymbol{\varphi}}{\partial x}\right)^2\right\}
+\frac{vm^2}{2g}\varphi^2_z, \label{nlsanisotropy}
\end{equation}
where 
\begin{equation}
m^2=\frac{2(\Delta-1)\left(-1+\alpha\right)}
{a^2\left(1-4\alpha\right)^{1/2}}.
\label{m2}
\end{equation}
We have obtained the Hamiltonian density of 
the anisotropic
non-linear $\sigma$ model.
The corresponding Euclidean (2-D classical) Lagrangian density is
\cite{Affleck1,Affleck2,Affleck-Les}
\begin{equation}
{\mathcal L}=\frac{1}{2 g}\left\{
\left(\partial_{\mu} \boldsymbol{\varphi} \right)^2+{m}^2
\varphi_z^2
+\frac{\theta}{8\pi}\epsilon^{\mu\nu}\boldsymbol{\varphi}\cdot
\left(
\partial_{\mu}\boldsymbol{\varphi}\times
\partial_{\nu}\boldsymbol{\varphi}
\right)
\right\}
\end{equation}
where we set $v=1$ for this discussion.
Since we take an integer spin case,
we can neglect the topological term.

Renormalization group equations of this model 
without the topological term are obtained by
Nelson and Pelcovits\cite{Nelson-Pelcovits}.
Since $d=2$ (dimension) and $n=3$ (number of components)
in our case,
we obtain the differential equations for
the ``dressed'' coupling constant $\tilde{g}$ and 
the ``dressed'' mass $\tilde{m}$:
\begin{eqnarray}
\frac{{\rm d}\tilde g}{{\rm d}\ln b}&=&\frac{1}{2\pi}
\left(\frac{\tilde g^2}{1+\tilde{m}^2}\right), \\
\frac{{\rm d}\tilde{m}^2}{{\rm d}\ln b}&=&2 \tilde{m}^2-\frac{1}{\pi}
\frac{\tilde{g}\tilde{m}^2}{1+\tilde{m}^2}.
\end{eqnarray}
Removing $\ln b$, we can obtain the renormalization relation
between $\tilde{m}^2$ and $\tilde{g}$,
\begin{equation}
\frac{{\rm d}\tilde{m}^2}{{\rm d}\tilde{g}}=
\frac{4\pi}{\tilde{g}^2}\tilde{m}^2\left(1+\tilde{m}^2\right)
-\frac{2\tilde{m}^2}{\tilde g}.
\label{RGeq}
\end{equation}
Therefore, the relation between the dressed 
$\tilde{m}^2$ and the dressed $\tilde{g}$ is
\begin{equation}
\tilde{m}^2(\tilde{g})\propto \exp\left(-\frac{4\pi}{\tilde g}\right).
\label{mass}
\end{equation}
For $\tilde{m}^2>0$, all flow lines terminate in a fixed line at $\tilde{m}^2=+\infty$.
One of these lines is believed to terminate at the BKT transition point
$\tilde{g}_{\textrm c}$.
Therefore,
this flow line should be a critical line
between the XY phase and the disordered (Haldane) phase.
From the similar discussion, we obtain 
the Ising critical line for $\tilde{m}^2<0$ case.
Finally,
in terms of the original Hamiltonian (eq. (\ref{anisotropicspin}))
by the use of eq. (\ref{m2}),
we now obtain the critical line eq.(\ref{nelpel1})
expressed as
\begin{equation}
\left|\Delta -1\right| \propto \frac{\sqrt{1-4\alpha}}{1-\alpha}
\exp{(-2\pi \sqrt{1-4\alpha})}.\label{nelpel}
\end{equation}
Note that eq. (\ref{nelpel}), or eq. (\ref{nelpel1}), 
can be extended to the $\alpha \to -\infty$ limit.

\end{document}